\documentclass[conference]{IEEEtran}
\IEEEoverridecommandlockouts
\usepackage{cite}
\usepackage{amsmath,amssymb,amsfonts}
\usepackage{algorithm}
\usepackage{algorithmic}
\usepackage{graphicx}
\usepackage{textcomp}
\usepackage{xcolor}
\usepackage{semantic}
\usepackage{float}
\usepackage{hyperref}

\newcommand{\tuple}[1]{\ensuremath{\left \langle #1 \right \rangle }}

\def\BibTeX{{\rm B\kern-.05em{\sc i\kern-.025em b}\kern-.08em
    T\kern-.1667em\lower.7ex\hbox{E}\kern-.125emX}}
\begin{document}

\title{Preserving Power Optimizations Across the High Level Synthesis of Distinct Application-Specific Circuits
}

\author{
\IEEEauthorblockN{Paulo Garcia}
\IEEEauthorblockA{\textit{International School of Engineering}, \textit{Chulalongkorn University}, Bangkok, Thailand \\
paulo.g@chula.ac.th}
}

\maketitle

\begin{abstract}
We evaluate the use of software interpretation to push High Level Synthesis of application-specific accelerators toward a higher level of abstraction. Our methodology is supported by a formal power consumption model that computes the power consumption of accelerator components, accurately predicting the power consumption on new designs from prior optimization estimations. We demonstrate how our approach simplifies the re-use of power optimizations across distinct designs, by leveraging the higher level of design abstraction, using two accelerators representative of the robotics domain, implemented through the Bambu High Level Synthesis tool. Results support the research hypothesis, achieving predictions accurate within $\pm$1\%. 
\end{abstract}


\section{Introduction}
Power efficiency has long been a desired property of custom digital circuits \cite{bhowmik2017power}. With the emergence of dark silicon \cite{esmaeilzadeh2011dark}, it is now a mandatory property. Decades ago, when the scale of custom circuits was much smaller (i.e., many solutions re-used programmable circuitry), it was feasible to spend a considerable portion of the design process on power optimizations \cite{allen2000custom}. However, as the deployment of custom circuitry has scaled up, especially due to the availability and efficiency of Field Programmable Gate Arrays (FPGA \cite{stewart2019verifying}), power optimizations must now be supported by practices embedded in the larger design methodology, much like has been done for other properties (e.g., determinism \cite{gomes2015rt}); ideally, at levels of abstraction above Register Transfer Level (RTL).
\par We evaluate the feasibility of using software interpreters, transformed into circuitry through the use of High Level Synthesis (HLS), where interpreted software represents application specific designs; pushing the design toward a higher level of abstraction. Optimizations can then be preserved across distinct solutions through compiler optimizations. I.e., we evaluate the research question: "Can an additional level of abstraction, in HLS design, enable re-usable power optimizations across distinct designs?". Specifically:

\begin{itemize}
    \item We describe a design methodology that preserves power optimizations across the design of distinct application-specific circuits, showing how software interpreters can be used, through HLS, to seamlessly implement application-specific circuits.
    \item We show how optimizations can be applied to the the interpretation engine, leveraging optimizations from one application-specific design toward another. We present a formal power model that can be used to estimate prior consumption \textit{a priori}.
    \item We evaluate all aspects of our approach using open-source an HLS tool, with FPGA vendor proprietary software for power analysis.  
\end{itemize}

\section{Related Work}

Our approach is inspired by Intermediate Fabrics, first introduced by Coole and Stitt \cite{coole2010intermediate}. Extant derivative techniques include software-defined FPGA acceleration \cite{lai2021programming}, more commonly referred to in the literature as FPGA virtualization \cite{vaishnav2018survey}. We point interested readers to \cite{quraishi2021survey} for a comprehensive survey on the topic.
\par Directly comparable to our work are HMFlow \cite{5771262} and RapidStream \cite{10.1145/3490422.3502361}. HMFlow introduced "hard macros":  previously synthesized and routed hardware pipelines that enable rapid design assembly. On the other hand, RapidStream focuses on optimizing synthesis time by co-optimizing C-to-RTL and RTL-to-bitstream compilation stages. In contrast, our approach takes advantage of recent advances in the C-to-RTL route to bring traditional software techniques to circuit design through new programming paradigms \cite{fryer2020towards} and, unlike related work that focuses on synthesis time, focusing on the re-use of power optimizations.

\section{The High Level Synthesis Model}

The model presented in this paper focuses on High Level Synthesis based on imperative programming languages; notably, the C/C++ family; this will be assumed explicitly from here onward. HLS based on other language semantics (e.g., functional or dataflow) must be modeled differently; we point readers to \cite{fryer2023good} for a review of HLS frameworks across programming paradigms.

\begin{figure}
\includegraphics[width=0.45\textwidth]{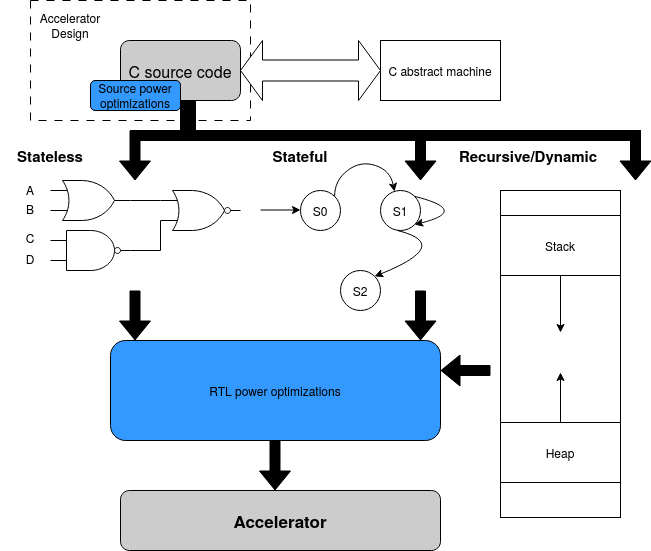}
\centering
\caption{Extant High Level Synthesis flow: power optimizations depicted in blue.}
\label{fig:hls}
\end{figure}

\subsection{Extant models}

Traditional HLS design converts the semantics of the abstract machine implied by the (software) language into hardware constructs. Stateless computations (i.e., using only arithmetic and logic without storage re-use) are trivially converted; stateful computations such as loop constructs (within function bodies) are typically converted into equivalent finite state machines that preserve sequential behavior, with calls to other functions implemented as sub-state machines. Challenges arise when abstract machine semantics do not have (simply) equivalent hardware constructions: infamously, some pointer arithmetic that implies global, shared access memory; runtime recursion that assumes unbounded stack space; and, dynamic allocation that assumes available heap space. All of these issues are well documented in the literature \cite{de2014high}; a conceptual view is depicted in Fig. \ref{fig:hls}.

\par Formally, we can notate HLS translation processes using operational reduction rules from the denotational semantics of the source language (in the case of C, using evolving algebras \cite{gurevich1992semantics}) to the denotational semantics of the target language (RTL), such that: 

\begin{algorithm}[H]
\begin{center}
$->$\inference[]{s}{h}
\end{center}
\end{algorithm}

denotes the translation of the software construction $s$ into the hardware construction $h$. Consider the following evolving algebra for a while loop, where \textit{"MoveTo"} is a macro that signifies the transfer of control from the current task (\textit{"CurrTask"}) to its argument.

\begin{algorithm}[H]
\begin{algorithmic}
\STATE Algebra for \textbf{"while"} construct:
\STATE $NextTask = CurrTask$
\STATE $MoveTo(Evaluate(TestValue(CurrTask)))$
\IF {$TestValue(CurrTask) = TRUE$} 
  \STATE $MoveTo(TrueTask(CurrTask))$
  \STATE $MoveTo(NextTask)$
\ELSE
    \STATE $break$
\ENDIF 
\end{algorithmic}
\end{algorithm}

\par For the sake of brevity, let us denote \textit{"MoveTo"} by $\rightarrow$, \textit{"TrueTask"} by $+$, \textit{"TestValue"} by $t$, and \textit{"Evaluate", "CurrTask" and "NextTask"} by $e$, $c$, and $n$, respectively. Using ternary if notation and commas to separate sequential statements, we can then denote the computation by:

\begin{algorithm}[H]
\begin{algorithmic}
\STATE $\textbf{while:} n = c, \rightarrow(e(t(c))), t(c) ? \rightarrow(+(c)), \rightarrow(n) : break$
\end{algorithmic}
\end{algorithm}

The equivalent hardware construct is a state machine with a set of states $S$ and a set of inputs $\Sigma = \{start, true, false, done\}$, with an initial state $s_0 \in S$ (idle), a condition evaluation state $s_1  \in S$ (which may be a sub-state machine, if the condition is not a stateless computation), a true task execution state $s_2  \in S$ (which is certainly a sub-state machine, at least prior to optimizations), and state transition function $\delta(S,\Sigma)$ such that $\delta(s_0,start) = s_1$, $\delta(s_1,true) = s_2$, $\delta(s_1,false) = s_0$, $\delta(s_2,done) = s_0$.  
\par Thus, we can define the HLS translation process for the while loop example as such:

\begin{algorithm}[H]
\begin{center}
$->$\inference[]{n = c, \rightarrow(e(t(c))), t(c) ? \rightarrow(+(c)), \rightarrow(n) : break}{S = \{s_0, s_1 = ->\inference[]{e(t(c))}{...}, s_2->\inference[]{+(c)}{...}\}, \Sigma , \delta(S,\Sigma) }
\end{center}
\end{algorithm}

where states $s_1$ and $s_2$ are recursively defined as the translations of the while loop condition and body, respectively, and output function is omitted for brevity. It is important to note that this transformation is injective; i.e., $->$\inference[]{s_1}{h_1}, $->$\inference[]{s_2}{h_2}, $s_1 \neq s_2 \longrightarrow h_1 \neq h_2$.


Notice that we do not attempt to give a full semantics of HLS here, which requires a mapping of every possible construct defined in the syntax of the source language: that is outside the scope of this paper, and best left to HLS tool implementers, but these examples should suffice to understand the rest of this paper.

\par Power optimization in traditional HLS is challenging, since it is not necessarily clear how changes to source code will be reflected in final hardware design, so re-factoring for power is a tool-specific iterative process. Furthermore, traditional hardware power strategies (clock and power gating) are not available at the semantic level of the HLS source language, and must be applied at either the RTL stage, or at the source stage through tool-specific \textit{pragmas} (compiler directives) and libraries.
\par Let $P(h)$ denote the power consumption (static and dynamic) of a hardware circuit $h$, and $s\prime$ denote source-level refactoring towards minimizing power consumption (e.g., by merging loops for state machine minimization \cite{dimitriou2016source}). Then, the expectation is:

\begin{algorithm}[H]
\begin{center}
$->$\inference[]{s_1}{h_1}, $->$\inference[]{s_1\prime}{h_1\prime} $\longrightarrow P(h_1) > P(h_1\prime)$
\end{center}
\end{algorithm}

Further optimizations at the RTL level are modeled as $\hat{h} = o(h)$, where $o(h)$ denotes the optimization function including, but not limited to, clock- and power-gating, refactoring, circuit minimization techniques, look-up optimization, etc.

\subsection{Our contribution: modeling higher level interpretation}

\begin{figure}
\includegraphics[width=0.45\textwidth]{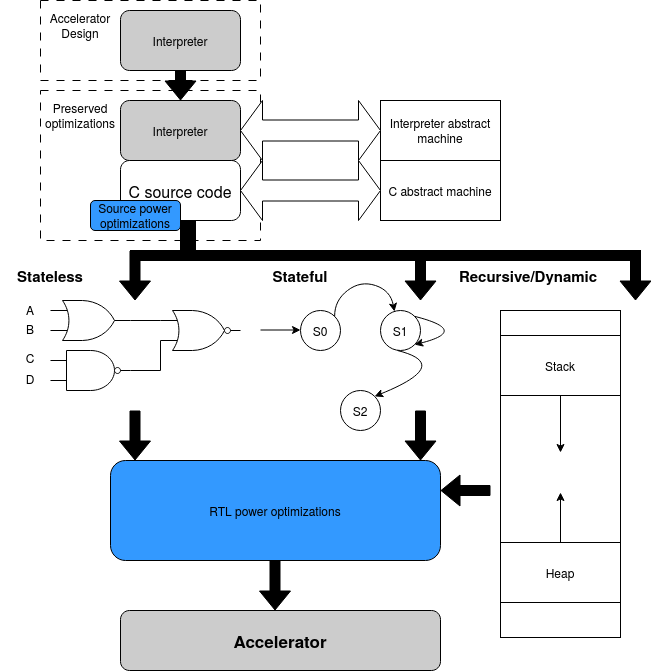}
\centering
\caption{This paper's contribution: moving the design towards a higher level of abstraction.}
\label{fig:ours}
\end{figure}

Our methodology lifts the level of abstraction, implementing interpretation (and subsequent compiler optimizations) as a means to design application-specific circuits. I.e., the final hardware design is a function of a computational engine, implemented at the level of traditional HLS source, interpreting behavior at a higher level of computation: from the programming languages' theory perspective, that of an embedded language. A conceptual view is depicted in Fig. \ref{fig:ours}.

\par The interpretation engine is a tuple given by $\tuple{E,P,F}$, where $E$ is a finite collection of storage elements $\tuple{e_0,e_1,e_2, ...}$ (where the size of each element may vary) that hold solution data, where $\tau(E)$ denotes the current state of the interpreter, i.e., the specific collection of values in $E$; $P$ is a stored program, consisting of an ordered sequence $(p_0, p_1, p_2, ...)$ of individual instructions; and $F$ is a collection of functions $\tuple{f_0(...), f_1(...), f_2(...), ...}$ where each function is of the form $f : \tuple{\tau(E),p} \rightarrow \tau \prime(E)$, i.e., a mapping of the current state of $E$ to a new state, given an instruction to execute. Notice that, while not necessary, functions in this model can be mapped to the hard macros introduced by HMFlow \cite{5771262}.
\par A specific solution is built by creating (or re-using) atomic functions and specifying a program that controls the flow of data across them. A function can be as simple as an arithmetic operation, or as complex as an encryptor/decryptor, with input instructions specifying operation to perform (such that $F$ is indexed by the current $p \in P$) and parameters to use. It is also expected that $P$ be cyclic, i.e., perform a non-terminating loop over the sequence of instructions, corresponding to the powered-on behavior of the final circuit (inner loops are, of course, allowed).
\par Consider a simple example, for a specific program $P_i = (p_{i_0}, p_{i_1}, ..., p_{i_n})$ and a specific collection of functions $F_i$. Because this is known at compile-time, the implemented software construction $s_i$ can be modeled as a while loop where the condition is always true, of the form:

\begin{algorithm}[H]
\begin{algorithmic}
\STATE \textbf{$s_i$:} $n = c, \rightarrow(+(c)), \rightarrow(n) $
\end{algorithmic}
\end{algorithm}

where the next task corresponds to the stored program interpretation; i.e.,

\begin{algorithm}[H]
\begin{algorithmic}
\STATE \textbf{$s_i$:} $n = c, \rightarrow( f_{p_{i_0}}, f_{p_{i_1}}, ... , f_{p_{i_n}}   ), \rightarrow(n) $
\end{algorithmic}
\end{algorithm}

\par Power optimizations can be applied once, to the interpretation engine, using the traditional HLS design flow. The advantage is that optimizations, achieved after long iterative processes, utilizing tool-specific methodologies, can now be preserved across distinct hardware designs, orthogonally to the functionality of the upper abstraction level. This promotes the re-usability of power optimizations across different projects, embedding the low-power concerns within the broader design flow. Note that the synthesis of each function $f_i \in F_i$ corresponds to a unique hardware circuit, assuming $E$ is immutable:  

\begin{algorithm}[H]
\begin{center}
$->$\inference[]{f_{p_{i_0}} = \tuple{\tau(E),p_{i_0}} \rightarrow \tau \prime(E)}{h^{f_{p_{i_0}}}}
\end{center}
\end{algorithm}

Thus, the synthesis of $s_i$ corresponds to:

\begin{algorithm}[H]
\begin{center}
$->$\inference[]{\mathbf{s_i:} n = c, \rightarrow( f_{p_{i_0}}, f_{p_{i_1}}, ... , f_{p_{i_n}}   ), \rightarrow(n)}{S = \{s_0, s_1 = h^{f_{p_{i_0}}}, s_2 = h^{f_{p_{i_1}}}, ...\}, \Sigma , \delta(S,\Sigma) }
\end{center}
\end{algorithm}

Source level optimizations to the implementation of functions result in the following optimized circuit (optimizations highlighted in blue for ease of legibility):

\begin{algorithm}[H]
\begin{center}
$->$\inference[]{\mathbf{s_i\prime:} n = c, \rightarrow( \textcolor{blue}{f\prime_{p_{i_0}}, f\prime_{p_{i_1}}}, ... , f_{p_{i_n}}   ), \rightarrow(n)}{S = \{s_0, \textcolor{blue}{s_1 = h\prime^{f_{p_{i_0}}}, s_2 = h\prime^{f_{p_{i_1}}}}, ...\}, \Sigma , \delta(S,\Sigma) }
\end{center}
\end{algorithm}

Further, RTL-level optimizations can be applied to functions, resulting in the following optimizations (further highlighted in green):

\begin{algorithm}[H]
\begin{center}
$->$\inference[]{\mathbf{s_i\prime:} n = c, \rightarrow( \textcolor{blue}{f\prime_{p_{i_0}}, f\prime_{p_{i_1}}}, ... , f_{p_{i_n}}   ), \rightarrow(n)}{S = \{s_0, \textcolor{blue}{s_1 = h\prime^{f_{p_{i_0}}}}, \textcolor{green}{s_2 = \hat{h}\prime^{f_{p_{i_1}}}}, ...\}, \Sigma , \delta(S,\Sigma) }
\end{center}
\end{algorithm}

\par Now, consider a new software construction $s_j$, corresponding to a new circuit design, that shares function use with $s_i$. It is then trivial to construct $s_j\prime$, as long as the condition:

\begin{algorithm}[H]
\begin{center}
$\exists f_p \in (s_i \cap  s_j), f\prime_p \in s_i\prime $
\end{center}
\end{algorithm}

holds; i.e., that we are re-using previously optimized functions.

\subsection{Our contribution: power estimation}

Approximations of expected power consumption for a given complete circuit can be obtained by linear combinations of prior estimations of functions' individual consumption, with accuracy constrained by the correctness of composition (routing, i.e., necessary additional gates, registers and wires to transfer data across different circuit areas) cost estimates. Further optimization at the RTL level can only improve power consumption, so modeling at this level should be considered a worst-case estimation.
\par Let $P(h)$ (recall: total power consumption of a circuit $h$) be decomposed into static and dynamic power, such that $P(h) = P_s(h) + P_d(h)$. Of course, $P_d(h)$ is excitation-specific: we use mean dynamic power as the working assumption. Because routing costs are not negligible, we must incorporate data transfer requirements. Given a square binary routing matrix $\mathbf{R}$:

\begin{equation}
    \mathbf{R} = \begin{bmatrix}
                    a_{1,1} & a_{1,2} & a_{1,3} & \dots  & a_{1,n} \\
                    a_{2,1} & a_{2,2} & a_{2,3} & \dots  & a_{2,n} \\
                    \vdots & \vdots & \vdots & \ddots & \vdots \\
                    a_{n,1} & a_{n,2} & a_{n,3} & \dots  & a_{n,n} 
                 \end{bmatrix}
\end{equation}

where index $a_{i,j} = 1$ represents the need for a connection from $f_i$ to $f_j$ (with $i=j$ representing input from top level), and the vector $\mathbf{D}$ be a column vector:

\begin{equation}
    \mathbf{D} = \begin{bmatrix}
                    d_{1} \\
                    d_{2} \\
                    \vdots \\
                    d_{n}  
                 \end{bmatrix}
\end{equation}

where index $d_i$ is the heuristic static cost of routing data to function $f_i$, as a proportion of data size. Given a circuit consisting of the composition of $n$ sequential functions:

\begin{algorithm}[H]
\begin{algorithmic}
\STATE \textbf{$s_i$:} $n = c, \rightarrow( f_{p_{i_0}}, f_{p_{i_1}}, ... , f_{p_{i_n}}   ), \rightarrow(n) $
\end{algorithmic}
\end{algorithm}

, then a static power approximation can be written as:

\begin{equation}
\begin{aligned}
    P_s(\mathbf{s_i}) = P_s(c) + \sum_{j=0}^{n-1} \left( P_s(f_{p_{i_j}}) \right) + \mathbf{1^T}(\mathbf{RD}) + w_1,\\ w_1\sim\mathcal{N}(0,\sigma ^{2})
\end{aligned}
\end{equation}

where $P_s(c)$ is a constant, corresponding to the static power of the circuitry required for the interpretation, including $E$ (which can be estimated \textit{a priori}); $1^T$ is the sum vector and $w_1$ is a random variable representing routing costs expected variability ($\sigma ^{2}$ can be empirically derived for distinct tools).

\par Given the assumption that only one function is active at any given time, i.e., there is no parallelism at the interpretation level (although there is no restriction on doing so in practice), average dynamic power is given by the average of individual functions' dynamic power. Let $l(P)$ denote the length of $P$. such that: 

\begin{equation}
\label{eq:dynamic}
\begin{aligned}
    P_d(\mathbf{s_i}) = P_d(c) + \Gamma + \mathbf{1^T}(\mathbf{RD}) + w_2 ,\\
    \Gamma= \frac{\sum\limits_{j=0}^{l(P)-1} \left(\alpha_{f_{p_{i_j}}}  P_{d}^{on}(f_{p_{i_j}}) + (1-\alpha_{f_{p_{i_j}}}) P_{d}^{off}(f_{p_{i_j}})    \right)}{l(P)},\\
    w_2\sim\mathcal{N}(0,\sigma ^{2})
\end{aligned}
\end{equation}

where $\alpha_{f_{p_{i_j}}}$ represents the proportion of total time per period that $f_{p_{i_j}}$ executes, and again $w_2$ is a random variable representing  (average) routing costs' expected variability. $P_d^{on}$ and $P_d^{off}$ represent average dynamic power when function is executing and idle (assuming no clock or power gating), respectively.

\section{Experiments and Results}\label{sec:results}

\begin{figure}
\includegraphics[width=0.5\textwidth]{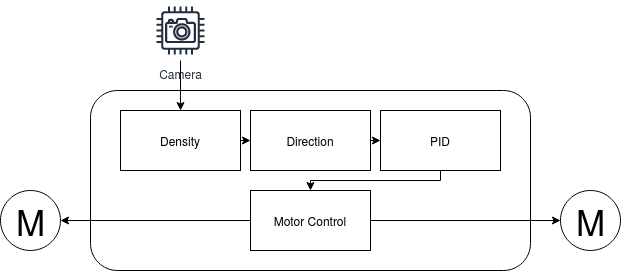}
\centering
\caption{"Chaser" block diagram.}
\label{fig:chaser}
\end{figure}
\begin{figure}
\includegraphics[width=0.4\textwidth]{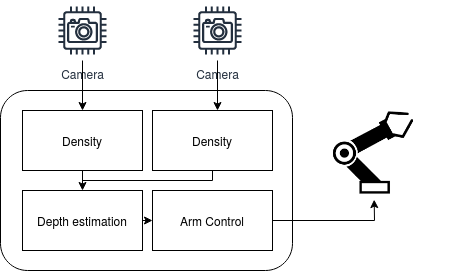}
\centering
\caption{"Grabber" block diagram.}
\label{fig:grabber}
\end{figure}

We evaluate our methodology using two example accelerator designs, representative of the robotics domain. 
\par Design 1 ("chaser") implements an object-following mobile robot. Chaser is equipped with an optical camera, used to determine the direction of the highest-density (visually) object, and 2 motors that effect movement, subject to a controller. All image processing and motor control is performed on chip (Fig. \ref{fig:chaser}). Design 2 ("grabber") implements a robotic arm. Grabber is equipped with a stereo-camera, used to determine the depth distance of the highest-density (visually) object, and a controller operating on a robotic arm that attempts to grab that object (Fig. \ref{fig:grabber}). Both designs share one function, implemented once in chaser and twice in grabber, and are otherwise distinct.
\par We evaluate optimizations using the Bambu HLS framework \cite{ferrandi2021bambu}, with power results further obtained from FPGA vendor tools, post-synthesis (in our case, Xilinx Vivado). Shared function is optimized through loop perforation (i.e., halving total number of pixels that must be traversed for object detection) and by arithmetic reduction (decreasing total number of computations required per pixel). We empirically estimate  power consumption for the baseline and optimized versions of chaser, and for the baseline version of grabber, and show that power consumption in the optimized version of grabber can be estimated (and optimizations are correctly preserved) by our model. All code and results are available in open-source form here\footnote{\url{https://github.com/paulo-chula/HIgher-abstraction-HLS.git}}.

\subsection{Formal description of prototypes}

Bambu provides post-HLS information about the total number of states implemented in the corresponding hardware output, allowing us to formally describe chaser as:

\begin{algorithm}[H]
\begin{center}
$->$\inference[]{\text{Chaser:} n = c, density, direction, pid, motors , \rightarrow(n) }{S_C, \lvert S_C \rvert =139, \Sigma , \delta(S,\Sigma) }
\end{center}
\end{algorithm}
    
with the optimized version denoted by:

\begin{algorithm}[H]
\begin{center}
$->$\inference[]{\text{Chaser:}\prime n = c, \textcolor{blue}{density\prime}, direction, pid, motors , \rightarrow(n) }{S_C, \lvert S_C \rvert =129 + \textcolor{blue}{10}, \Sigma , \delta(S,\Sigma) }
\end{center}
\end{algorithm}

\par In both cases, routing and costs matrices $R_C$ and $D_C$ (expressed as routing power of number of bits) correspond to:

\begin{equation}\label{eq:matrix_c}
    \mathbf{R_C} = \begin{bmatrix}
                    1 & 1 & 0 & 0 \\
                    0 & 0 & 1 & 0 \\
                    0 & 0 & 0 & 1 \\
                    0 & 0 & 0 & 0 
                 \end{bmatrix}, 
    \mathbf{D_C} = \begin{bmatrix}
                    P_r(32) \\
                    P_r(64) \\
                    P_r(128) \\
                    P_r(64)  
                 \end{bmatrix}
\end{equation}

\par Similarly, grabber is defined as:

\begin{algorithm}[H]
\begin{center}
$->$\inference[]{\text{Grabber:} n = c, density, density, depth, motors , \rightarrow(n) }{S_G, \lvert S_G \rvert =33, \Sigma , \delta(S,\Sigma) }
\end{center}
\end{algorithm}
    
with the optimized version denoted by:

\begin{algorithm}[H]
\begin{center}
$->$\inference[]{\text{Grabber:}\prime n = c, \textcolor{blue}{density\prime}, \textcolor{blue}{density\prime}, depth, motors , \rightarrow(n) }{S_G, \lvert S_G \rvert =13 + \textcolor{blue}{20}, \Sigma , \delta(S,\Sigma) }
\end{center}
\end{algorithm}

with corresponding matrices $R_G$ and $D_G$:

\begin{equation}
    \mathbf{R_G} = \begin{bmatrix}
                    1 & 1 & 0 \\
                    0 & 0 & 1  \\
                    0 & 0 & 0  
                 \end{bmatrix}, 
    \mathbf{D_G} = \begin{bmatrix}
                    P_r(32) \\
                    P_r(32) \\
                    P_r(128) 
                 \end{bmatrix}
\end{equation}

\subsection{Results}

\begin{table}[]
\centering
\begin{tabular}{l c c c}
Prototype & Logic Slices & FFs\\
\hline
    Chaser         & 38087   &5122 \\
    Chaser$\prime$ & 37661   &4813\\
    Grabber        & 9036  & 3294 \\
    Grabber$\prime$& 8587  & 2752
 \end{tabular}
\caption{Prototypes' FPGA resource usage.}
\label{tab:area}
\end{table}

\begin{table}[]
\centering
\begin{tabular}{l c c}
Prototype & Dynamic & Static\\
\hline
    Chaser         &  5.895 & 0.095\\
    Chaser$\prime$ &  5.023 & 0.092\\
\hline
\hline
Function & Dynamic (on) & Dynamic (off)\\
\hline
Density         &  6.486 & 2.162\\
Density$\prime$         &  5.598 & 1.942\\
Direction & 1.124 & 0.924\\
PID & 0.712 & 0.698\\
Motor control & 2.295 & 1.133\\
 \end{tabular}
\caption{Chaser empirical power analysis, before and after optimizations, including individual functions' on and off dynamic power. All results in Watts.}
\label{tab:power_chaser}
\end{table}

Table \ref{tab:area} depicts hardware utilization required for the un-optimized and optimized versions of both prototypes. Power results for chaser, before and after optimization, are depicted in Table \ref{tab:power_chaser}.
\par We can now apply our model to estimate the benefits of power optimization and the costs of routing, predict what power savings will be achieved on grabber, and compare with experimental results. For sufficiently large, or sufficiently large number of, functions, $P_d(c) \ll P_d(\mathbf{s_i})$, so we can approximate and simplify Equation \ref{eq:dynamic} by assuming $P_d(c) = 0$. $l(P)$ can be approximated by assuming equivalence to the number of executed states, i.e., the cardinality of $S$ ($\lvert S \rvert$), where $\alpha_{f_{p_{i_j}}}$ then becomes the number of states in the transformation

\begin{algorithm}[H]
\begin{center}
$->$\inference[]{f_{p_{i_j}} = \tuple{\tau(E),p_{i_j}} \rightarrow \tau \prime(E)}{\lvert S \rvert \in h^{f_{p_{i_0}}}}
\end{center}
\end{algorithm}

, which are given by Bambu. Thus, plugging the numbers for chaser from Table \ref{tab:power_chaser} and matrices in Equation \ref{eq:matrix_c} into Equation \ref{eq:dynamic}, we obtain:

\begin{equation}
\label{eq:dynamic_app}
\begin{aligned}
    P_d(\text{Chaser}) = \Gamma + \mathbf{1^T}(\mathbf{R_CD_C}) + w_2 ,\\
    \Gamma= \frac{\sum\limits_{j=0}^{l(P)-1} \left(\alpha_{f_{p_{i_j}}}  P_{d}^{on}(f_{p_{i_j}}) + (1-\alpha_{f_{p_{i_j}}}) P_{d}^{off}(f_{p_{i_j}})    \right)}{\lvert S_C \rvert} \equiv\\
    \equiv 5.895 = \Gamma + P_r(288) + w_2, \\
     \Gamma= \frac{ 721.312   }{139} = 5.227, \equiv \\
     \equiv  P_r(288) = 0.668 + w_2 \equiv\\
     \equiv P_r(1) = 0.0023 + w_3,\\
    w_2\sim\mathcal{N}(0,\sigma ^{2})\\
    w_3\sim\mathcal{N}(0,\sigma ^{2} (\frac{1}{288})^2)
\end{aligned}
\end{equation}

, allowing us to obtain a high confidence estimation for routing costs per bit, and already knowing the power decrease due to the approximation of function \textit{density}. Plugging this estimation back into Equation \ref{eq:dynamic} using corresponding values for grabber (baseline and optimized versions) results in predictions $P_d(\text{Grabber}) = 7.498 +  w_3, w_3\sim\mathcal{N}(0,\sigma ^{2}(\frac{1}{288})^2)$ and  $P_d(\text{Grabber}\prime) = 5.937 +  w_3, w_3\sim\mathcal{N}(0,\sigma ^{2}(\frac{1}{288})^2)$ , compared to empirical results depicted in Table \ref{tab:power_grabber} which report $P_d(\text{Grabber}) = 7.505$ and $P_d(\text{Grabber}\prime) = 5.898$.

\begin{table}[]
\centering
\begin{tabular}{l c c}
Prototype & Dynamic & Static\\
\hline
    Grabber         &  7.505 & 0.099\\
    Grabber$\prime$ &  5.898 & 0.095\\
\hline
\hline
Function & Dynamic (on) & Dynamic (off)\\
\hline
Density         &  6.486 & 2.162\\
Density$\prime$         &  5.598 & 1.942\\
Depth & 0.322 & 0.064\\
Motor control & 0.428 & 0.102\\
 \end{tabular}
\caption{Grabber power estimation, compared to empirical power analysis, before and after optimizations, including individual functions' on and off dynamic power. All results in Watts.}
\label{tab:power_grabber}
\end{table}

\section{Conclusions}

We described an implementation strategy for the abstraction of HLS through interpretation that allows preserving power optimizations across distinct designs. Alongside it, we described a formal model of HLS translation, accompanied by a power consumption model that can accurately estimate the impact of power optimizations on one design, based on their impact in prior designs. Experiments on two different accelerators representative of the robotics domain support the validity of our contributions.
\par Future work must explore model extensions to account for parallelization. Our assumption is based on sequential execution of accelerated functions. Whilst this is the easiest way to perform C-to-RTL translation, it does not necessarily take advantage of parallelization opportunities. Doing so, and preserving model capabilities, will require extensions that allow us to consider different combinations of concurrent functions.

\bibliographystyle{IEEEtran}
\bibliography{IEEEabrv,refs}

\end{document}